\documentstyle[seceq]{ptptex}

\input epsf.sty

\newcommand{\deffig}[4]{
  \begin{figure}[t]
    \begin{center}
      \ \null \epsfxsize=#3 \epsfbox{#2}
      \caption{#4}
      \label{fg:#1}
    \end{center}
  \end{figure}
}
\newcommand{\Fig}[1]{
  Fig.\ref{fg:#1}
}

\title{Optimization Algorithms Based on Renormalization Group}

\subtitle{Application to 2D Edwards-Anderson Spin Glass}

\author{
Naoki {\sc Kawashima}\footnote{E-mail address: nao@phys.metro-u.ac.jp}
}

\inst{
Department of Physics, Tokyo Metropolitan University,
Minami-Ohsawa 1-1, Hachiohji, Tokyo 192-0397
}

\recdate{
}

\abst{
Global changes of states are 
of crucial importance in optimization algorithms.
We review some heuristic algorithms in which global updates 
are realized by a sort of real-space 
renormalization group transformation.
Emphasis is on the relationship between the structure of
low-energy excitations and ``block-spins'' appearing in the algorithms.
We also discuss the implication of existence of
a finite-temperature phase transition
on the computational complexity of the ground-state problem.
}

\begin{document}

\maketitle

\section{Introduction}
In the last twenty years or so, we have realized that the
interaction between physics and information sciences, such as
operations research, optimization, pattern recognition and 
theories of learning, can be useful in studies of all of these fields.
For example, the problem of spin glasses\cite{EdwardsA1975,SherringtonK1975} 
has been providing us with important insights not only for solid 
state physics but also for artificial intelligence\cite{Hopfield1982}.
In addition, Kirkpatrick, Gelatt and Vecci\cite{KirkpatrickGV1983} 
demonstrated that the knowledge of statistical mechanics
could be used in optimization problems by introducing fictitious 
temperature into those problems, which opened up possibilities 
of applying or generalizing various physical ideas to other fields.

Study of physics has also benefited from the information sciences.
One of the most important unanswered questions 
in physics of disordered systems is the one concerning the 
nature of the low temperature phase in spin glasses in three dimensions.
While few analytical means is available for answering this question,
the numerical approach also has a serious difficulty.
Namely, when one uses the Monte Carlo method, which is the most commonly
used numerical technique for disordered systems, 
one always encounters a severe slowing down which makes it practically
impossible to study a sizable system below the critical temperature.
This difficulty seems to be closely related to the fact that the
problem of finding the ground states of, say, the Edwards-Anderson (EA)
spin glass model in three dimension, is NP-hard\cite{Barahona1982}.
In general, therefore, if one develops an efficient heuristic algorithm for
NP-hard optimization problems, the above-mentioned difficulty
in study of spin glasses would be also removed.
In fact, such an attempt was made by Gr\"otschel et al\cite{GroetschelJR1985},
who tried to apply the linear programming technique to the general
spin glass optimization problems.

It is suggestive to compare the case of two dimensions 
with that of three dimensions.
In two dimensions, the ground state problem is not
NP-hard and some polynomial time algorithms were
proposed\cite{BiecheMRU1980,BarahonaMRU1982}.
However, in this case, the system does not have a phase transition
at any finite temperature.
In addition, many heuristic algorithms have been shown to be
very efficient in two dimensions while in three dimensions
few of them has turned out to be powerful for large scale problems.

In this paper, we present another example of application of a 
physical idea to optimization problems.
While the method discussed below works more efficiently than some other
naive optimization methods such as the simulated annealing,
in three dimensions it works not as efficiently as in two dimensions,
similar to other heuristic methods.
It is, however, illuminating to take a close look at this method
because for this method we can clearly see the reason 
for relatively poor performance in higher dimensions and
it casts a new light on the relationship between physical 
nature of a system and computational complexity of its ground
state problem.

Renormalization group is one of most important ideas that
the physics of this century produced.
We attempted to construct a heuristic optimization algorithm
exploiting this idea\cite{KawashimaS1992}.
There, we proposed a renormalization transformation
in which multiple initial solutions
are used to decompose the whole system into small pieces, i.e.,
``block spins''\cite{Kadanoff1976}.
Then, neglecting all internal degrees of freedom inside each piece,
we obtain a new ``renormalized'' problem.
The method was successfully applied to the spin glass problem in
two dimensions.
For models in higher dimensions, however, it was not clear if
the algorithm was as efficient or not, because 
we could not clearly see the asymptotic behavior of the
required computational time as a function of the problem size.
Later, Usami and Kano\cite{UsamiK1995} used the renormalization group 
idea for optimization in a different approach.
They claimed that the method works fine for the traveling
salesman problem.
Recently, Houdayer and Martin\cite{HoudayerM1999} combined
the above mentioned renormalization transformation
with the genetic algorithm.
They demonstrated that the resulting algorithm is as efficient
as other ``state-of-art'' heuristic optimization algorithms for
a wide class of problems including spin glasses in three dimensions and
the traveling salesman problem.

In this paper, we concentrate on the EA spin glass model
described by the following Hamiltonian for concreteness:
$$
  {\cal H} = - \sum_{ij} J_{ij} S_i S_j
$$
where $S_i$ being an Ising variable and $J_{ij}$ a quenched random variable.

\section{Cross Breeding Operation}

\deffig{Renormalization}{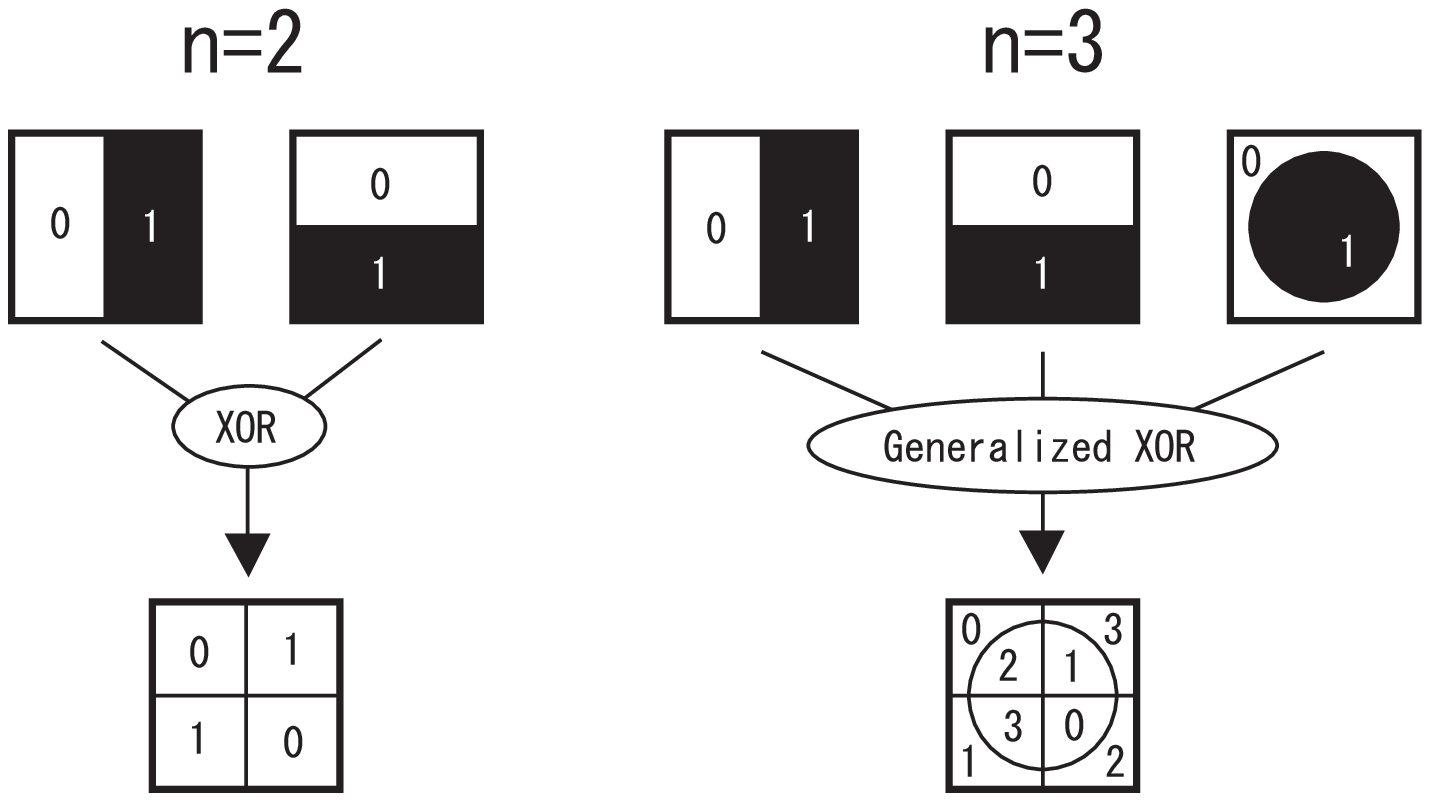}{80mm}
{The renormalization (or block-spin) transformation.
In diagrams in the upper row, black patches marked with ``1'' 
represent regions where spin orientations are oppsite to those
in a reference ground state whereas white ones with ``0'' stand for
regions with the same spin orientations as the reference state.
}
Our first attempt\cite{KawashimaS1992} 
to combine the renormalization group idea with
heuristic optimization algorithms was
inspired by Swendsen and Wang's replica Monte Carlo method\cite{SwendsenW1986}
for simulating spin glass systems at finite temperature.
Swendsen and Wang considered an ensemble of identical systems, which they
called ``replicas'', each at different temperature.
By comparing two replicas at slightly different temperatures,
they decomposed into clusters of spins.
To be more specific, they considered the exclusive-or of 
two Ising spins, each locating at the corresponding site 
on a different replica.
Regarding the resulting 0 and 1 as two ``colors'',
a two-toned map of the system was obtained. (See the ``$n=2$''
case in \Fig{Renormalization}.)
A cluster of spins, then, was defined as a group of spins with the
same color connected by couplings $J_{ij}$.
We can choose the inter-cluster coupling constants so that the
detailed balance condition in terms of the original spins
may be satisfied even in the updating process where clusters,
rather than the individual original spins, are used as updating units.
In this way, Swendsen and Wang succeeded in reducing the 
autocorrelation time by orders of magnitude in two dimensions.
The success was, however, not as complete for three dimensions as 
in the two dimensional case.

One of the obvious reasons why their algorithm does not work
in higher dimensions is that the clusters grow too big.\cite{ChaosLength}
In the system at a certain temperature, regions of the size of 
the correlation length are randomly activated.
It would be desirable to identify each of such
activated clusters.
If we compare only two systems at the same temperature, however,
two adjacent regions are often ``accidentally'' activated in both
systems.
In such cases, these two clusters are not recognized as separate objects.
Since we have only two colors in Swendsen and Wang's algorithm, 
it is increasingly probable that the same color is assigned to
two sizable regions adjacent to each other in higher dimensions.
With a small number of colors, therefore,
structures of a size comparable to the correlation length are
not effectively updated.

In order to overcome this difficulty we considered the following
cross-breeding operation\cite{KawashimaS1992}.
The inputs to this operation are multiple ``parents'',
i.e., spin configurations,
and the outcome is an ``offspring'', a configuration
which is better than, or at least as good as, the best among
the parents.
The number of parents, $n$, is adjusted to maximize the efficiency
of the whole algorithm.
A cluster, or a ``block spin'', is defined as 
the maximal connected set of spins in which an arbitrary two spins 
have the same relative orientation for all the parents.
Technically speaking, we assign a ``color'' defined by
$$
  c_i \equiv \sum_{\mu=2}^n 2^{\mu-2} \left(\frac{S^{(1)}_i S^{(\mu)}_i +1}{2}
  \right)
$$
to each site $i$,
where $S^{(\mu)}_i$ is the $i$-th spin in the $\mu$-th parent.
This is a natural generalization of the exclusive-or.
We, then, define a cluster as a connected set of sites 
with the same color.
Once clusters are identified, we regard each cluster as a single
renormalized spin and assign a one-bit degree of freedom to it.

The next step is quenching all $n$ configurations
with these renormalized spins as updating units.
Relative orientation of original ``bare'' spins inside each renormalized spin 
does not change hereafter throughout the whole cross-breeding operation.
After this quench procedure, we compare resulting configurations and
eliminate duplication, i.e., we discard a configuration if there is another
one identical to it.

Next, we again compare the resulting configurations 
and redefine block spins in the same fashion as described above.
New block spins consist of previous block-spins and therefore larger
than those.
We ``heat up'' all the remaining parents but the one 
with the lowest energy.
As a result, these configurations become random in terms of block spins.
We then quench all remaining configurations and eliminate duplication.

These procedures are repeated until the only one configuration remains.
Finally we take the last configuration as the ``offspring''.
Since we do not heat up the configuration with the lowest energy at any stage,
it is guaranteed that the energy of the offspring is not
larger than the lowest energy among parents.


\section{Various Structures of Algorithms}
\deffig{Structures}{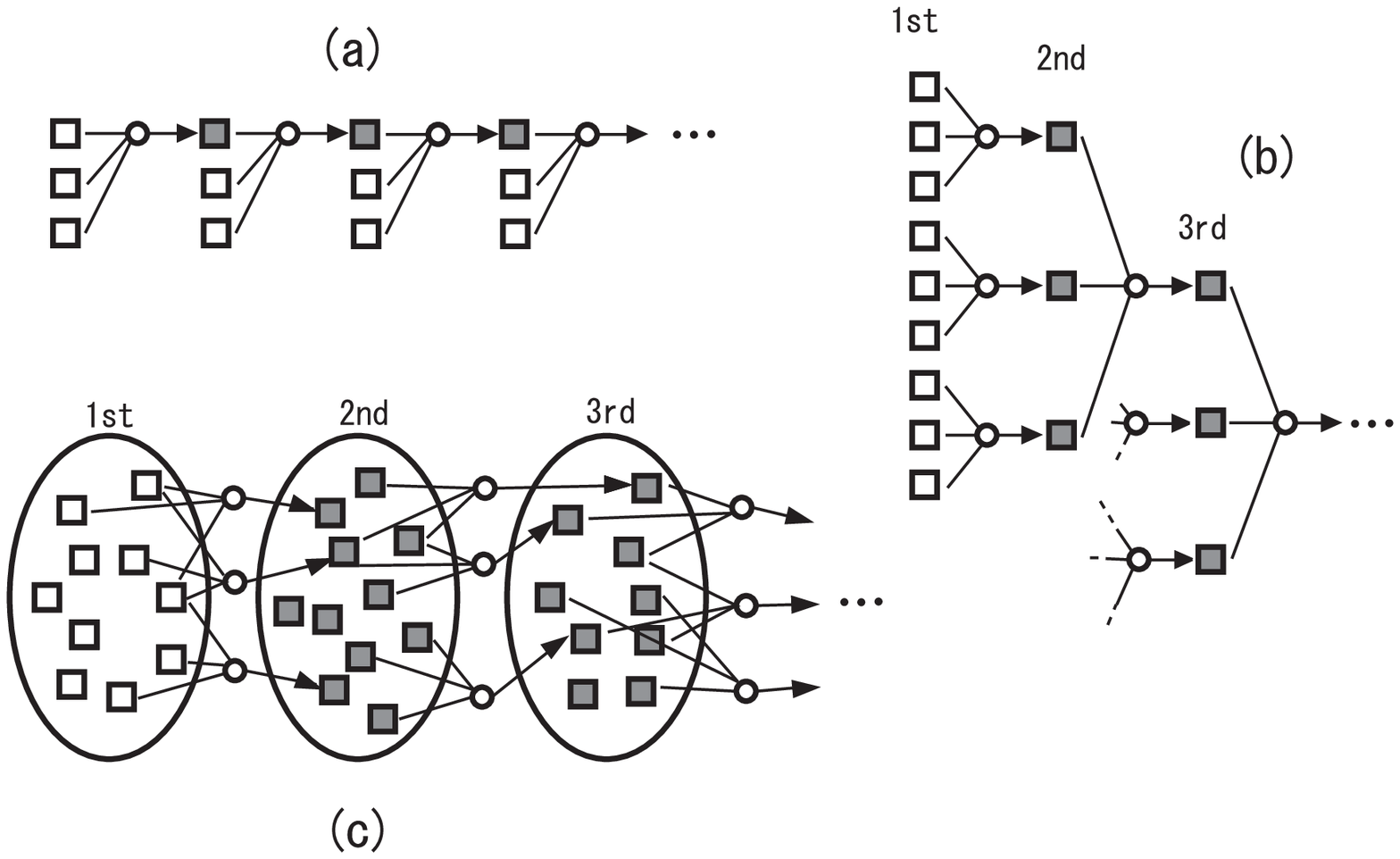}{120mm}
{Three different structures of algorithms that consist of
``cross-breeding'' operations. The chain structure (a),
the tree structure (b), and the genetic pool structure (c).
Open circles represent the cross-breeding operations whereas
open and filled squares stand for, respectively, spin configurations
of the first generation and those of higher generations.}
Besides the difficulty due to too large clusters,
there is yet another point that we have to consider
in order to cast Swendsen and Wang's idea into a form of an
optimization algorithm.
Namely, in their algorithm the block spin transformation
is repeatedly applied only to the
configurations at a certain fixed temperature.
While this was sufficient for a finite temperature Monte Carlo simulation,
it would be certainly insufficient for optimization because
the cluster size will never grow large.
The fixed cluster size is as problematic as too big clusters,
because such fixed-size clusters can not update the system effectively
at low temperatures where the correlation length is much larger than
the cluster size.

\subsection{Chain Structure}

In the replica optimization method\cite{KawashimaS1992},
we applied the cross-breeding operation described above repeatedly 
to the outcome of the previous cross-breeding operation,
expecting that as a configuration goes through more cross-breeding operations
the size of block spins becomes larger until finally it becomes
as large as the whole system.
There, we proposed two options for implementing this idea.
One is repeated application of the cross-breeding along an ``assembly line''
((a) in \Fig{Structures}) and the other is the cross-breeding
operations organized as a hierarchical tree ((b) in \Fig{Structures})
as we discuss below.
In the ``assembly-line'' structure, at each cross-breeding,
one of the inputs is the outcome of the last cross-breeding 
whereas the other inputs are spin configurations of the first generation.
The spin configurations of the first generation
are created from random spin configurations through quenching.
Here quenching is nothing but the poor technique of optimization
often called the ``greedy algorithm''.
Therefore, in general, there is a large difference in the fitness of
the best input and those of the others.
As a result in most cases the offspring resembles the best input
much more than the others.
The other parents work as perturbations to the currently-best solution.

The advantage of this structure is the easiness of implementation.
In addition, the computational cost for moving to the next generation 
does not increase as we go to higher generations.
On the other hand, we cannot expect to gain from one application of 
the cross-breeding in this structure as much as we can get from 
going to the next generation in the other structures discussed below.
However, for small scale applications, we found that this structure
was advantageous.

\subsection{Hierarchical Tree Structure}
While the assembly-line structure is simple to implement,
we expect that it does not find ground states efficiently
in larger scale applications.
As the first parent becomes better, it is increasingly difficult 
to find large clusters by which the first parent
can be transformed into better states, 
only with perturbations produced by the first generation 
configurations.
This is because usually the block spins
found in cross breeding operation with the first generation 
configurations are small.
Therefore it is desirable to choose equally good $n$ spin configurations
as the inputs of the cross-breeding operations.
This naturally leads to the tree structure.

Namely, we consider a hierarchical tree with a leaf attached 
to each end point at the tree top, i.e.,
the leftmost column in (b) of \Fig{Structures}.
The procedure goes rightward from the tree top.
At each branching point, represented by an open circle in \Fig{Structures},
we apply the cross-breeding operation to the parents.
This results in a new spin configuration which is, then,
placed at the position next to the branching point.
We repeat this procedure for all branching points 
until we get the last leaf for the root, the
right-most one.

As we go rightwards on the tree, 
the typical size of renormalized spins becomes larger.
In two dimensions, the exponent, $\theta$, which characterizes
the dependence of the excitation energy upon the size of spin clusters 
is negative.
Therefore as we proceed in the algorithm, the typical 
excitation energy we handle becomes smaller.
This is why the ``smaller-scale-first-larger-scale-later''
strategy works fine at least in two dimensions.
This may mean that the present strategy works only for systems
that does not have a phase transition at a finite temperature,
since the negative value of $\theta$ signifies
the absence of a phase transition at a finite temperature.
For general cases where larger scale structures may correspond to
larger energy, a more sophisticated consideration will be required.

Using this method, we have studied two-dimensional spin glass systems
to find that the thermal exponent does not agree with the stiffness
exponent\cite{KawashimaReview1999,KawashimaS1992,KawashimaHS1992}
in contrast to what had been generally taken as granted.
We also found recently\cite{Kawashima1999} that the elementary excitations
in this model are fractal clusters of spins and that the scaling exponent
characterizing their excitation energy agrees with the thermal exponent
whereas it certainly differs from the stiffness exponent.

\subsection{Genetic-Pool Structure --- Genetic Algorithm}
Houdayer and Martin\cite{HoudayerM1999} proposed yet another structure
for the algorithm.
Namely, they combined the cross-breeding operation with the 
genetic algorithm\cite{GeneticAlgorithm}.
The genetic pool of the first generation is an ensemble of
the first-generation spin configurations. (See (c) in \Fig{Structures}.)
In order to generate a configuration of the $(g+1)$-th generation,
they choose $n$ configurations randomly from the pool of the $g$-th generation,
cross-breed them, and put the offspring into the $(g+1)$-th
generation pool.
They claimed that the resulting algorithm is efficient not only
for the spin glass problems in two dimensions but also for a wider class
of optimization problems, including the three-dimensional EA model
and the traveling salesman problem, although the systematic study
of the asymptotic efficiency of the algorithm has not been carried out.

\section{A Complementary Algorithm}
Usami and Kano\cite{UsamiK1995} proposed another method for optimization
inspired by the renormalization group idea.
The method is specialized for the traveling salesman problem.
They considered decompositions of systems into cells of the same
size and shape.
They first consider such a decomposition on a large scale and
obtain a coarse-grained problem.
Then, using the obtained global approximate solution, 
they consider another decomposition on a smaller scale.
Their approach is complementary to ours in that
their method deals with large scales first and 
small scales later.
They claimed that the method is very efficient for the traveling
salesman problem.
In light of the relationship between the length scales and the
energy scales discussed above, 
this ``larger-scale-first'' strategy may be appropriate
because in a typical instance of the traveling salesman problem,
larger length scale correspond to larger energy, i.e., $\theta > 0$.

\section{Discussions and Summary}
We have reviewed a few heuristic optimization algorithms based on
the idea of the renormalization group.
Since their asymptotic efficiency has not been thoroughly investigated,
it is not clear if these approaches are useful for larger scale problems.
However, at least, they show very clear advantage over other
heuristic algorithms such as the simulated annealing for various
optimization problems of interest.
We have also discussed the relationship among phase transitions,
the sign of the excitation-energy exponent $\theta$,
and computational complexity of problems.
Our discussion suggests that the simultaneous optimization 
on different scales may be required for a wider class of problems.

\section*{Acknowledgements}
This work is supported by Grant-in-Aid for Scientific Research Program
(No.11740232) from Mombusho, Japan.


\begin{thebibliography}{99}
\bibitem{ChaosLength}
  This ``correlation length'' may be affected if the difference in the
  temperature is larger than some crossover value because of the
  so-called chaos effect in temperature.
\end{thebibliography}
\end{document}